\documentclass[preprint]{aastex}
\def\be{\begin{equation}}
\def\ee{\end{equation}}
\def\HI{\ion{H}{1}}
\def\HII{\ion{H}{2}}
\def\HA{H$\alpha$}
\begin{document}
\title{Star Formation Around Supergiant Shells in the LMC}
%

\author{Laura G.\ Book\altaffilmark{1,2,3}, You-Hua Chu\altaffilmark{2}, Robert 
A.\ Gruendl\altaffilmark{2}, Yasuo Fukui\altaffilmark{4}}
\altaffiltext{1}{\itshape Department of Physics,
 University of Illinois at Urbana-Champaign, 1110 West Green Street, Urbana, 
IL 61801}
\altaffiltext{2}{\itshape Department of Astronomy, University of Illinois 
at Urbana-Champaign, 1002 West Green Street, Urbana, IL 61801}
\altaffiltext{3}{\itshape Currently at Department of Physics, California 
Institute of Technology, Pasadena, CA 91125}
\altaffiltext{4}{\itshape Department of Physics and Astrophysics, 
Nagoya University, Chikusa-ku, Nagoya 464-8602, Japan}

%
%
\begin{abstract}

We examine the recent star formation associated with four 
supergiant shells (SGSs) in the Large Magellanic Cloud (LMC): 
LMC 1, 4, 5, and 6, which have been shown to have simple 
expanding-shell structures. 
\HII\ regions and OB associations  are used to infer star 
formation in the last few Myr, while massive young stellar 
objects (YSOs) reveal the current ongoing star formation.
Distributions of ionized, \HI, and molecular components
of the interstellar gas are compared with the sites of 
recent and current star formation to determine whether 
triggering has taken place.  We find that a great majority 
of the current star formation has occurred in gravitationally 
unstable regions, and that evidence of triggered star 
formation is prevalent at both large and local scales.

\end{abstract}
\subjectheadings{ISM: general, bubbles, kinematics and dynamics;
stars: formation; Magellanic Clouds}

\maketitle
\pagestyle{empty}
%
%
\section{Introduction}  \label{sec:intro}

The interstellar medium (ISM) of late-type galaxies often exhibits
prominent, large shell structures with sizes approaching 1 kpc.  These
``supergiant shells'' (SGSs) are thought to be formed by the fast stellar
winds and supernova explosions of multiple generations of massive star
formation; they require 10$^{52} - 10^{53}$ ergs for their creation, the
equivalent of tens to hundreds of supernova explosions \citep{M80}.  
The expansion of a SGS can shock and sweep up interstellar gas, altering 
the physical conditions and distribution of the ISM.  Most notably, a 
SGS may puncture its galactic gas disk and vent hot gas into the
galactic halo, while its expansion in the galactic disk can compress 
ambient gas and induce star formation.

SGSs have been identified from \HA\ images of ionized gas or 21-cm
line data of neutral \ion{H}{1} gas.  These two types of surveys do not
lead to the same set of objects in a galaxy.  In Paper I (Book et al.\
2008), we used objects in the Large Magellanic Cloud (LMC) to illustrate
that not all \HA-identified SGSs are physical shells and that the
\HA-identified sample \citep{M80} and \ion{H}{1}-identified sample 
\citep{Kim99} differ mainly in their recent star formation history.  
In this paper, we investigate the role played by SGSs in star formation.

Star formation can be triggered by the expansion of a SGS via two 
methods \citep{Elm98}.  As a SGS expands, it sweeps up and collects the
ambient ISM in its shell, and when the dense shell gas cools, it may
become gravitationally unstable and collapse to form stars. 
Alternatively, the expansion of a SGS may compress preexistent 
dense clouds around its edge,  causing these clouds to collapse.  
Due to the proximity of the LMC \citep[50 kpc;][]{Feast99}, the SGSs in the
LMC provide ideal sites to observe star formation associated with SGSs and 
determine their star formation mechanism.  

\citet{Yetal01a,Yetal01b} have examined the distribution of
their 168 molecular clouds and young ($<$30 Myr) stellar clusters 
\citep{Betal96} with respect to the nine \HA-identified SGSs
in the LMC.  They find that the surface density of molecular clouds 
near the rims of SGSs is enhanced by a factor of 1.5--2 in both 
number and mass, while the surface density of the clusters inside the
SGSs is four times as high as that outside.  Furthermore, they find 
that 70\% of the clusters associated with molecular clouds along the
SGS rims are located on the interior-facing side of the clouds.
These results suggest that SGSs do play a significant role in both
the formation of molecular clouds and the dynamical triggering of
cluster formation.

We have investigated the most recent star formation in four 
\HA-identified SGSs in the LMC, LMC 1, 4, 5, and 6, which have 
been demonstrated in Paper I to be physical shells with a simple
expanding structure.
We use two observational indicators of recent star formation, \HII\ regions
and young stellar objects (YSOs).  \HII\ regions are photoionized by 
the UV radiation from hot massive stars, which are most frequently
found in OB associations.  Limited by the lifetime of O-type stars,
\HII\ regions trace the star formation that occurred within the last 
few Myrs.  Massive stars are formed rapidly; thus, massive YSOs trace 
the current, ongoing star formation within $<$1 Myr.  As YSOs are still
enshrouded, they have not significantly altered their ambient ISM,
and thus allow us to examine the interstellar conditions that have 
led to their formation.  Using \HII\ regions and massive YSOs, we 
are able to infer the recent star formation history of the SGSs and
further assess whether star formation was triggered in these areas.

This paper reports our findings on the star formation around these four
SGSs in the LMC.  In Section \ref{sec:obs} we describe the observations
upon which we base our analysis. In Section \ref{sec:sf} we compare
the distributions of OB associations, YSOs, and interstellar gas 
for each of the four SGSs, and discuss the implications for star
formation.  In Section \ref{sec:causes} we show that enhanced 
star formation along the SGS rims is statistically significant and
that star formation occurs mostly in gravitationally unstable regions.
Our conclusions are summarized in Section \ref{sec:conc}.

\section{Observations}  \label{sec:obs}

We have used \HA\ images of the SGSs to show the distribution of 
ionized gas, \HI\ observations to show the complete shell structure,
{\it Spitzer Space Telescope} infra-red (IR) observations to examine
the dust distribution and YSOs, and CO maps to show the distribution
of molecular gas in these SGSs.

\subsection{Optical \HA\ Images}  \label{sec:Ha}

We have used \HA\ images from the Magellanic Cloud Emission-Line Survey 
\citep[MCELS,][]{MCELS}, which was made with a CCD camera on
the Curtis Schmidt Telescope at Cerro Tololo Inter-American Observatory.
The images have an angular resolution of $\sim$3$''$, and were taken 
with a narrow-band interference filter centered on the \HA\ line 
($\lambda_{c} = 6563$ \AA, $\Delta\lambda=30$ \AA). The detector was 
a front-illuminated STIS 2048 $\times$ 2048 CCD with 21 $\mu$m pixels, 
giving a field of view of 1\fdg1 $\times$ 1\fdg1.  The \HA\ images were
mosaicked to cover the central 8$^{\circ}$ $\times$ 8$^{\circ}$ of the 
LMC.  From the mosaic we extracted regions covering the neighborhoods 
of the \HA-selected SGSs.

\subsection{\HI\ Observations}  \label{sec:HI}

The neutral hydrogen data were synthesized by \citet{Kim03}, combining
observations made with the Australia Telescope Compact Array (ATCA) 
and the single-dish Parkes Telescope. These observations cover 11\fdg1 
$\times$ 12\fdg4 of the sky, and the final synthesized images have an 
angular resolution of 1\farcm0 and a pixel size of $20''$. The 
observing band was centered on 1.419 GHz, corresponding to a central
heliocentric velocity of 297 km~s$^{-1}$.  The bandwidth used was 4 MHz, 
giving a complete velocity range of $-$33 to +627 km~s$^{-1}$ 
\citep{Kim98,Kim03}.

\subsection{Spitzer Space Telescope Observations}  \label{sec:SAGE}

The LMC has been surveyed by the {\it Spitzer Space Telescope} using 
both the InfraRed Array Camera \citep[IRAC;][]{Rieke04} and the 
Multiband Imaging Photometer for {\it Spitzer} \citep[MIPS;][]{Fazio04}
under the Legacy Program ``SAGE'' \citep{Meixner06}.  The observations
were made in the IRAC 3.6, 4.5, 5.8 and 8.0 $\mu$m and MIPS 24, 70, 
and 160 $\mu$m bands in 2005 July and October--November.  The survey
covers 7$^\circ \times 7^\circ$ on the sky, with a 7 $\times$ 7 array 
of 1\fdg1 $\times$ 1\fdg1 tiles of IRAC exposures and a 19 $\times$ 2 
array of 4$^\circ \times$ 0\fdg4 MIPS fastscans.  The angular 
resolutions are $\sim$2$''$ for the IRAC bands and $\sim$6$''$ for the 
MIPS 24 $\mu$m band, corresponding to 0.5 and 1.5 pc at the distance
of the LMC (50 kpc; Feast 1999).  The MIPS 70 and 160 $\mu$m images, 
having much lower resolutions, are not used in this study.

This paper uses mosaicked {\it Spitzer} images of SGSs made with the
archival SAGE data, and the YSO list made by \citet{GC08} using 
the SAGE data.  The masses of YSOs are uncertain, but 
in general those brighter than 8 mag at 8.0 $\mu$m represent massive
stars with masses greater than $\sim$10 $M_\odot$ and the fainter ones 
represent intermediate-mass stars with masses of $\sim$4--10 $M_\odot$
\citep{Cetal08}.
This YSO list does not include low-mass YSOs as they cannot be 
distinguished from background galaxies in the diagnostic 
color-magnitude diagrams.

\subsection{CO Data}  \label{sec:CO}

To compare the distributions of YSOs and molecular clouds around the
SGSs, we use the CO 2.6 mm line observations from the LMC survey 
made with the 4-m NANTEN telescope at Las Campanas Observatory in Chile
\citep{F99,Fukui01}.   The telescope has a beam size of 2\farcm6 and a 
pointing accuracy of 20$''$.  Observations were made every 2$'$ spacing
in position switching, covering 6$^{\circ} \times 6^{\circ}$ on the sky 
in a total of 32800 observed positions.  Note that the detection 
limit of the NANTEN survey was a few $\times 10^4$ $M_\odot$, thus
dust globules and small molecular clouds would not be detected.

\section{Star Formation Around SGSs}  \label{sec:sf}

Massive stars may dynamically alter the physical conditions 
of the ambient ISM via fast stellar winds and supernova 
explosions, and photoionize the ISM with UV radiation.
SGSs have dynamic ages greater than the lifetime of a
massive star; thus, the massive stars that were responsible 
for shaping the ISM must have been formed earlier than
those that are currently photoionizing the ISM.
The prolonged history of star formation and the interaction
between massive stars and the ISM in SGSs are best illustrated
by \HA\ images. 

In Figures 1-3, we present \HA\ images of SGSs LMC 1, 4, 5, and 6.
The filamentary morphology and shell structures amply manifest 
the dynamical effects of fast stellar winds and supernova 
explosions.  In these figures, we further plot the massive and 
intermediate-mass YSOs identified by \citet{GC08} in red and 
green circles, respectively, and uncertain intermediate-mass YSOs 
in green crosses.  The relative locations of YSOs and ionized gas
not only reveal the star formation history from a few Myr ago to 
present, but also allow us to assess whether the current star 
formation was triggered by stars from a previous generation
through the expansion of the SGSs.

The expansion of the SGSs may be responsible for two types of 
triggered star formation.   The first, collect and collapse, 
describes a situation in which the expansion of a shell causes
gas to collect along its rim, and the eventual cooling leads 
to fragmentation and gravitational collapse to form stars. The 
second mechanism involves the compression of preexisting clouds
by the expansion of a SGS to trigger star formation. 
These two mechanisms can be observationally discerned from the
distribution of YSOs and the interstellar gas.

We use \HI, neutral hydrogen, to trace the bulk gas and CO to 
trace dense molecular clouds.  In the top middle panels of 
Figures 1-3, we compare the distributions of \HI\ (in grey scale)
and CO (in contours) in the SGS shells and in their surroundings 
in order to assess whether the molecular clouds are compressed 
preexisting clouds or formed in the shell from cooled swept-up 
interstellar gas.  The {\it Spitzer} 8 $\mu$m images, in the top 
right panels of Figures 1--3, show good correlation with \HI\ 
images but at a much higher angular resolution.  

Below we examine the recent star formation around the SGSs
LMC 1, 4, 5, and 6 individually.  We compare the YSO locations 
with the distributions of OB associations and \HII\ regions 
to examine the star formation history in the last few Myr, 
and use the \HI\ and molecular components of the shell gas 
to assess the physical conditions of the environment of current star 
formation.  To avoid crowding, the \HII\ regions are marked in 
Figs.~1--3 of this paper, and OB associations marked in the
\HA\ images in Figs.\ 1, 4, and 5 of Paper I.

\subsection{LMC 1}

\begin{figure}[h!]
\epsscale{0.6}
\plotone{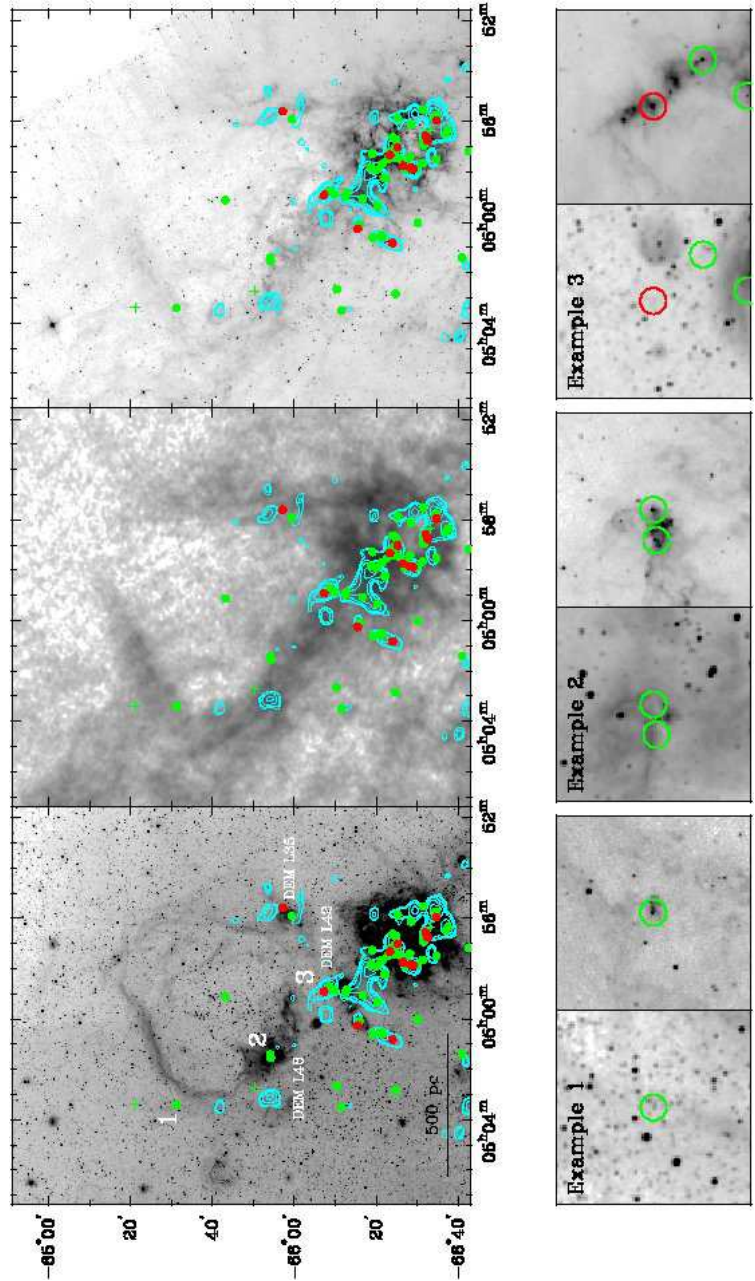}
\caption{The supergiant shell LMC 1.  Top row (left to right) - 
MCELS \HA\ image, ACTA+Parkes \HI\ column density map, and
{\it Spitzer} 8 $\mu$m image, superposed by NANTEN CO contours
(in cyan) and marked with locations of YSOs.  The massive YSOs 
in red circles, intermediate-mass YSOs in green circles, and 
uncertain intermediate-mass YSO candidates in green crosses.
The bottom row shows three examples of YSOs with pairs of \HA\
(left) and 8 $\mu$m (right) images. The locations of these
examples are marked in the \HA\ image in the top row.}
\label{fig:LMC1}
\end{figure}

LMC 1 has the simplest and most coherent shell structure among all 
SGSs in the LMC.  Its \HA\ image in Figure~\ref{fig:LMC1}
({\it top left}) shows the bright \HII\ region DEM\,L48 \citep{DEM76} 
at the southeast shell rim, a fainter \HII\ region DEM\,L35 at 
the southwest shell rim, and curved filaments delineating the 
rest of the shell.  The OB association LH16 \citep{lh} stretches 
from the center of LMC 1 to the bright southeast \HII\
region, possibly indicating a propagation of star formation.  
While the stars at the southeast end of LH16 are younger and 
responsible for the photoionization of the \HII\ region DEM\,L48, 
the stars at the northwest end of LH16 are older and are likely 
responsible for the formation of LMC 1's large shell structure.  YSOs 
associated with LMC 1 are mostly distributed around its periphery;
they are either inside \HII\ regions DEM\,L35 and 48 or near the 
\HII\ region DEM\,L42 at the outer southern boundary of LMC 1.
This distribution is consistent with the aforementioned conjecture 
that star formation has proceeded outwards in LMC 1. Away from these
\HII\ regions, only a low-level of star formation is observed, as one 
YSO lies along a wisp just outside the northeast rim of the
optical shell and another is projected toward the shell center.

The molecular clouds in LMC 1 fall mostly in dense regions along the 
\HI\ shell, but there is no good correlation between YSOs and molecular 
clouds.  Only one cloud near the \HII\ region DEM\,L42 contains YSOs;
the other less-massive clouds along the \HI\ shell are almost entirely
without YSOs.  If these YSO-less clouds originated from preexisting 
molecular clouds that had been swept into the expanding SGS shell, 
the compression would have been conducive to gravitational collapse and 
star formation.  The absence of bright YSOs indicates that these clouds
were probably formed from the gas collected in the SGS shell, but have
not collapsed yet.  It is also possible that these clouds are forming
only low-mass YSOs, which cannot be identified from {\it Spitzer} data
alone; high-resolution and sensitive images in the near-infrared are 
needed to search for low-mass pre-main-sequence stars.

The YSOs in the LMC 1 shell all fall near the inner rim of its
\HI\ shell.  The YSOs along the south rim of LMC 1 are all associated
with \HII\ regions that are photoionized by massive stars born in
the last few Myr.  The formation of these YSOs may have been affected
more directly by the local \HII\ regions than the SGS itself. 
Three examples of local environments of YSOs are given in the bottom
panels of Figure 1.  For each example, a pair of MCELS H$\alpha$ 
and {\it Spitzer} 8 $\mu$m images are displayed; the former shows 
the distribution of ionized interstellar gas and the latter shows 
YSOs (point sources) and warm dust and PAH emission (diffuse sources).
In Example 1, the YSO is located in the northeast quadrant of LMC 1 
at the inner boundary of the \HI\ shell; the expansion of the ionized
gas shell into the neutral shell may have triggered the formation of
this YSO.  In Example 2, the YSOs are in the \HII\ region DEM\,L48,
but the close-up H$\alpha$ image in the bottom middle panel of Figure 1
shows that the YSOs are located in dust lanes bordering ionized gas;
it is likely that the expansion of the \HII\ region triggered the 
YSO's formation.  In Example 3, the YSO is inside a giant molecular 
cloud along the south rim of LMC 1; no obvious triggering mechanisms
can be concluded.

We note that the YSO candidate (045907.4$-$654313.3) near the center 
of the LMC 1 shell may be a Be-like star with circumstellar material.
This object shows bright photospheric emission with $V  = 13.297 \pm 
0.094$, $(U-B)= -0.845 \pm 0.122$, and $(B-V)= -0.143 \pm 0.098$ 
\citep{Zetal04}, consistent with a B0-2\,III star with a reddening 
of $E(B-V)$ = 0.1--0.15.  Such post-main sequence early-type B stars 
are $\sim10^7$ yr old, similar to the age of the SGS LMC 1.  This YSO
candidate near the center of LMC 1 is probably not a bona fide YSO.

To the south of LMC 1 is the active star forming region N11, and
the \HI\ image suggests that N11 and LMC 1 may be physically connected.
It is unlikely that a causal relationship in star formation exists
between LMC 1 and N11, based on an examination of stellar ages in N11.
At the center of N11 is a superbubble blown by the 4--5 Myr old OB 
association LH 9, and the superbubble is surrounded by the $\sim$2 Myr
old OB associations LH 10 to the north and LH 13 to the east \citep{WP92}.
Clearly, the star formation in N11 started at its center and
propagated outwards.  The center of N11 is more than 200 pc from 
the rim of LMC 1.  There is no feasible physical mechanism to link
the star formation activities of N11 and LMC 1.

\subsection{LMC 4}

\begin{figure}[h!]
\epsscale{0.6}
\plotone{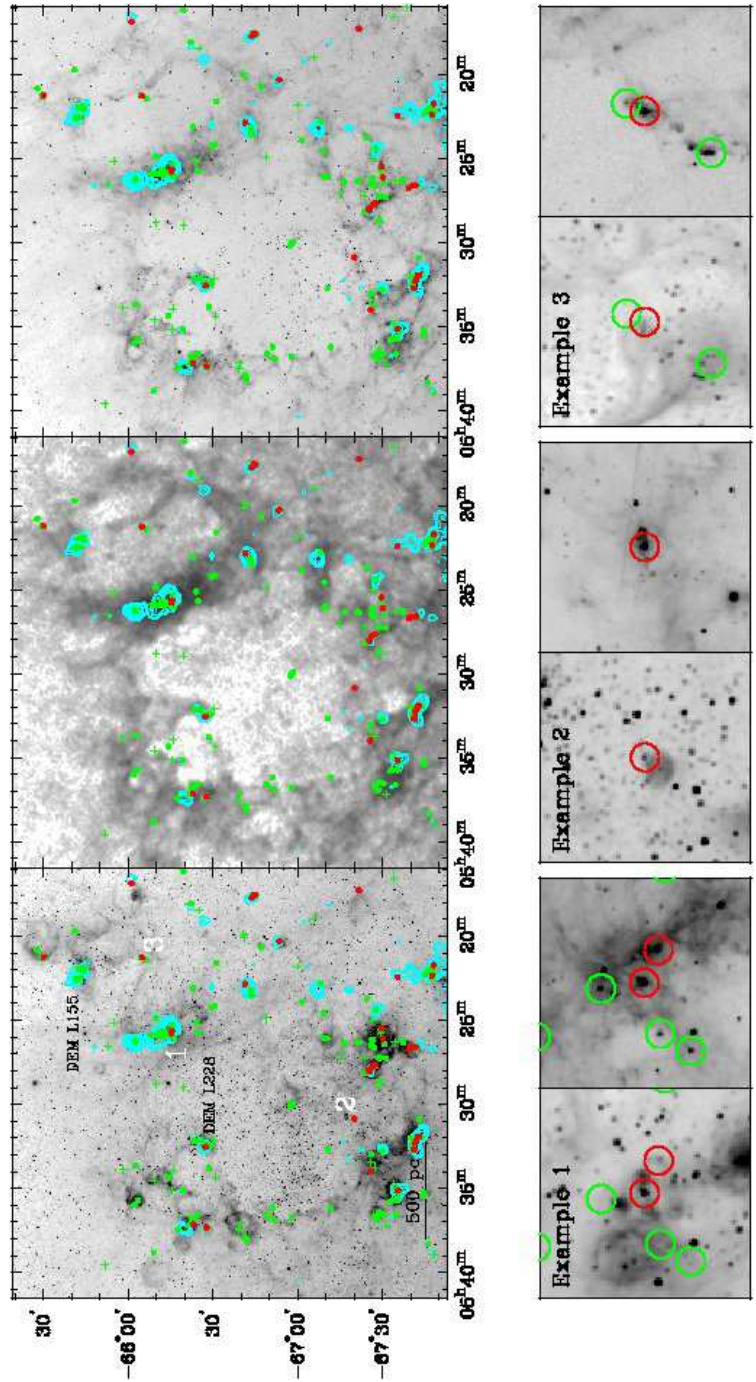}
\caption{Same as Figure 1, but for supergiant shells LMC 4 and 5.}
\label{fig:LMC45}
\end{figure}

LMC 4 is the largest SGS in the LMC, and has a roughly circular shape 
with a diameter of $\sim$ 1.2 kpc.  Its \HA\ shell is composed of many 
long filaments connecting a series of \HII\ regions. As noted in Paper I,
multiple OB associations exist in LMC 4: LH77 near the center of the shell
has an \HA\ arc near its western end, and LH72 is in an \HII\ region that extends
from the northern rim toward the shell interior, while LH51, 56, 57, 58, 
83, 91, and 95 are in \HII\ regions along the shell rim, and LH52 and 53
are located in \HII\ regions in the interaction zone between LMC 4 and LMC 5.
Apparently, star formation has proceeded outward in LMC 4.  
Figure~\ref{fig:LMC45} ({\it top left}) shows that most, $\sim$80\%, 
of the YSOs are associated with the \HII\ regions along the periphery, 
confirming the continuing star formation activity.  It also shows a
few YSOs within the central cavity of LMC 4, but usually superposed 
on diffuse \HA\ emission.

LMC 4 has a number of molecular clouds strung along its rim.
Considering the large volume of gas that has been swept into 
the shell, it is likely that the formation of molecular clouds 
is due to collect and collapse. The only exception is the cloud 
associated with LH72, which appears to be a preexisting cloud 
that has been compressed by the expansion of the LMC 4 shell. 
The concentration of molecular gas along the shell rim is most 
noticeable in the region between LMC 4 and LMC 5, where it appears 
that the collision of the two shells has further compressed the
shell gas and caused a very dense, large molecular cloud to form.

The most massive molecular clouds in LMC 4 are associated with 
\HII\ regions and OB associations.  YSOs are found in every 
molecular cloud along the rim of LMC 4, including the one coincident
with LH72 and DEM\,L228 in the northern extension into the shell 
interior.  The YSOs that are not associated with large molecular 
clouds are all in ionized gas, with a range of surface brightness
from prominent superbubbles to diffuse field gas.  These ionized 
interstellar structures require the injection of stellar energy, 
in the form of stellar winds and supernova explosions; thus their 
underlying stellar population must be older than those that are still
in dense \HII\ regions associated with molecular clouds.  The majority
of the YSOs not in molecular clouds are also located along LMC 4's
shell rim, similar to those in molecular clouds.  Overall, we see 
that most YSOs in the LMC 4 region are distributed along the shell 
rim and are in environments with a history of recent star formation.

LMC 4 has long been cited as a classic example of a large interstellar
shell consisting of swept-up material that has subsequently cooled,
fragmented, and collapsed to form stars \citep{Dom95}.
The distributions of molecular clouds, OB associations, \HII\ regions,
and YSOs in the LMC 4 shell are certainly consistent with this
assertion \citep{Yetal01a}.  At a detailed level, shocks and 
photoionization induced
implosions may be responsible for the onset of star formation.
Example 1 in Figure 2 shows active star formation along the
collision zone between LMC 4 and LMC 5; the compression may be
responsible for forming the giant molecular clouds, but the 
locations of YSOs at the boundary between ionized gas and
dust clouds indicate that local triggering still plays an
important role.  Example 2 shows a YSO interior to LMC 4; the
H$\alpha$ image reveals that the YSO is at the edge of a small
\HII\ region ($\sim$12 pc in diameter), which might have triggered
the YSO formation.

\subsection{LMC 5}

LMC 5 is located just to the northwest of LMC 4, and the collision of 
their shells has caused a density enhancement on the eastern side
of LMC 5.  \HII\ regions exist along the interaction zone between LMC 5
and LMC 4 and at the northern tip of LMC 5.  Three OB associations lie
along the rim of LMC 5, LH52 and 53 along the interaction zone and LH45
in the northern \HII\ region DEM\,L155.  No OB association is 
identified in LMC 5's shell interior.  Long \HA\ filaments on the
southern and western sides complete the shell of LMC 5.  Most YSOs in 
LMC 5 exist in the northern \HII\ region and the \HII\ regions along
the interaction zone.  Only three YSOs are located in the interior of 
LMC 5, and they are all associated with interior \HA\ filaments.

LMC 5 has three distinct molecular clouds along its rim.  The most
massive one is located in the interaction region between LMC 4 and 5,
superposed on the OB associations LH52 and 53 and their numerous 
\HII\ regions.  The second is in the north with the OB association 
LH45 and its \HII\ region DEM\,L155.  The third cloud is the least
massive one and sits at the south end of the interaction zone without
any noticeable \HII\ region.  Since there are small molecular clouds 
in the surroundings of LMC 5, it is not clear whether the molecular 
clouds around the shell rim were formed by collect and collapse or 
compression of preexisting clouds.  It is likely that both mechanisms
have taken place.

About 2/3 of the YSOs in LMC 5 are in these three molecular clouds.
The majority of the remaining 1/3 are located between the two molecular
clouds along the interaction zone and are associated with diffuse ionized 
gas, where the supernova remnant 0524-664 in DEM\,L175 is located.  
Only 2 or 3 YSOs are in diffuse ionized gas projected in the shell interior.  
We conclude that, similar to LMC 4, the expansion of LMC 5 may be
responsible for triggering the formation of OB associations, especially
in the collision zone against LMC 4, but local shocks and photoionization
triggered the formation of isolated stars.  For instance, Example 3 in
Figure 2 shows YSOs at the boundary between ionized gas and a long \HI\ 
filament inside LMC 5.

\subsection{LMC 6}

\begin{figure}[h!]
\epsscale{0.6}
\plotone{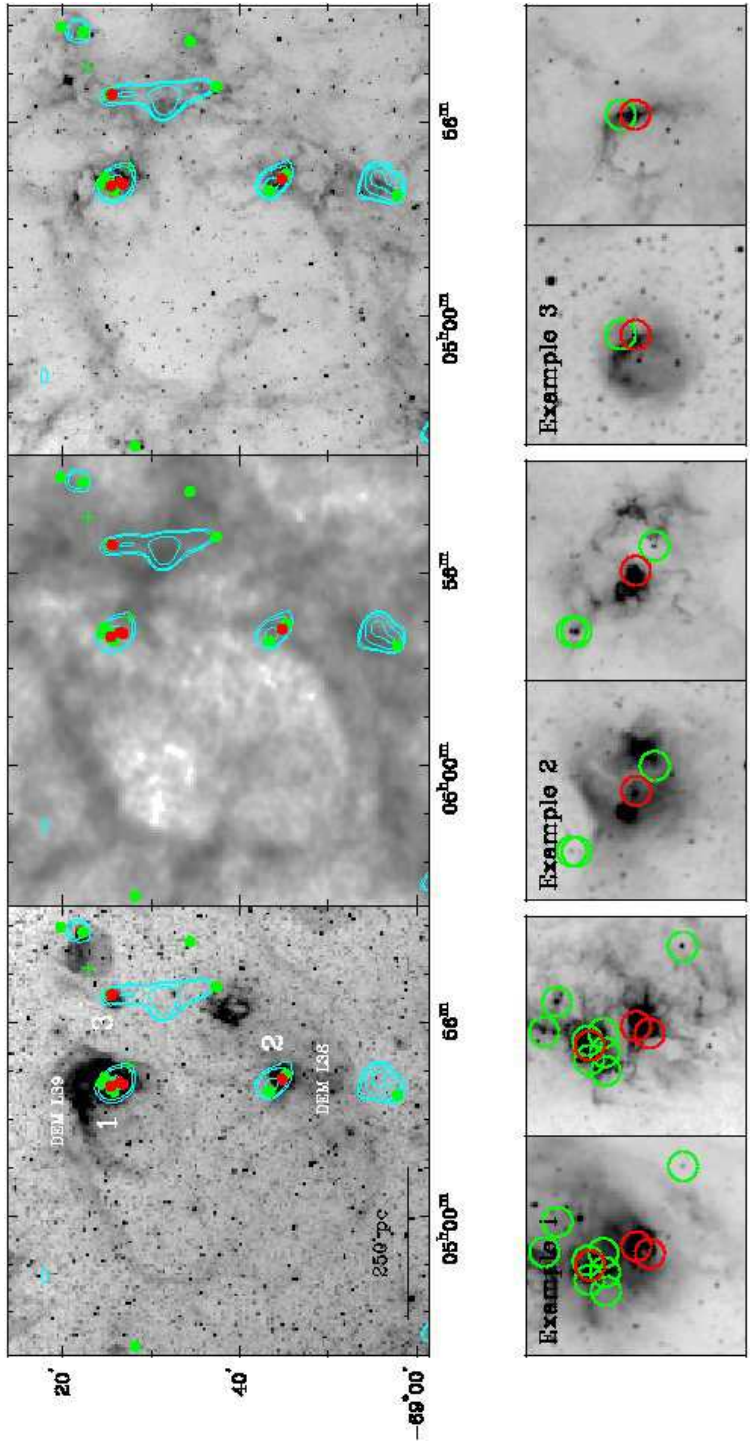}
\caption{Same as Figure 1, but for supergiant shell LMC 6.}
\label{fig:LMC6}
\end{figure}

LMC 6 is the smallest of the simple \HA-selected SGSs in the LMC. 
Figure~\ref{fig:LMC6} ({\it top left}) shows two \HII\ regions, 
DEM\,L38 and 39, along the western periphery of LMC 6.  
Long \HA\ filaments define the eastern half of the LMC 6 shell.  
Two OB associations, LH11 and 12, reside in DEM\,L38 and 39, 
respectively, but no OB associations have been identified within 
the LMC 6 shell interior, possibly owing to a lower concentration 
of former star formation.  It shows very little on-going star 
formation.  Only a few YSOs are present and all of them cluster within 
the \HII\ regions associated with the two OB associations. 

LMC 6 shows only two molecular clouds along its rim, associated with 
the OB associations LH11 and 12. Since these clouds are only on one 
side of the shell and there are also exterior molecular clouds, they
are likely preexisting clouds that have been simply compressed by the 
expansion of LMC 6.  The absence of detectable molecular clouds along
other parts of the shell rim may be attributed to the low density of 
the environment of LMC 6 and the small size of the shell so that not
enough gas mass has been swept into a shell to form large molecular 
clouds.

The star formation of LMC 6 contrasts with all of the previous shells. 
Few YSOs surround this SGS, and all fall within the dense molecular 
clouds that have formed the OB associations LH11 and LH12.  Two 
molecular clouds to the west, exterior of LMC 6, are also forming 
stars, but at a lower level than the two clouds along the rim of
LMC 6.  Compression by the expanding LMC 6 shell may not have
started the initial star formation in the two molecular clouds
along its rim, but must have helped to raise the star formation 
activity.

The formation of YSOs in the \HII\ regions DEM\,L38 and 39 appears
to be dictated by the local interstellar conditions.  Example 1 in 
Figure 3 shows the active star formation associated with the OB 
association LH12 in DEM\,L39; the environment is sufficiently 
complex that star formation triggers cannot be unambiguously
identified.  Example 2 shows the star formation in LH11 in 
DEM\,L38; some YSOs are located near boundaries between ionized
gas and dark clouds, while others are projected against the bright
emission of the \HII\ region.  Example 3 shows YSOs in a molecular
cloud outside LMC 6; again, YSOs are located near the boundaries
between ionized gas and dark clouds.

\section{Discussion}  \label{sec:causes}

\subsection{Statistical Analysis of the Star Formation Distribution} 
\label{sec:dist}

The distribution of YSOs with respect to the SGSs in Figures 1-3 appear
to show that most YSOs in the vicinity of each SGS are on the rim.
In order to quantitatively establish this we have measured the surface 
density of YSOs in annular regions that roughly match the shape of the 
\ion{H}{1} shell and extend from the SGSs interior, through the shell
rim, and around the immediate periphery of the shell.  
Specifically we use elliptical annuli that roughly match the 
\ion{H}{1} shells of LMC\,4, 5, and 6, and a series of trapezoidal 
annuli to approximate the shape of LMC\,1.
The trapezoidal annuli for LMC\,1 are truncated on the southern
side to avoid including sources in the N\,11 \ion{H}{2} complex and 
the average radius of each region is used for plotting purposes.

\begin{figure}[h!]
\epsscale{0.85}
\plotone{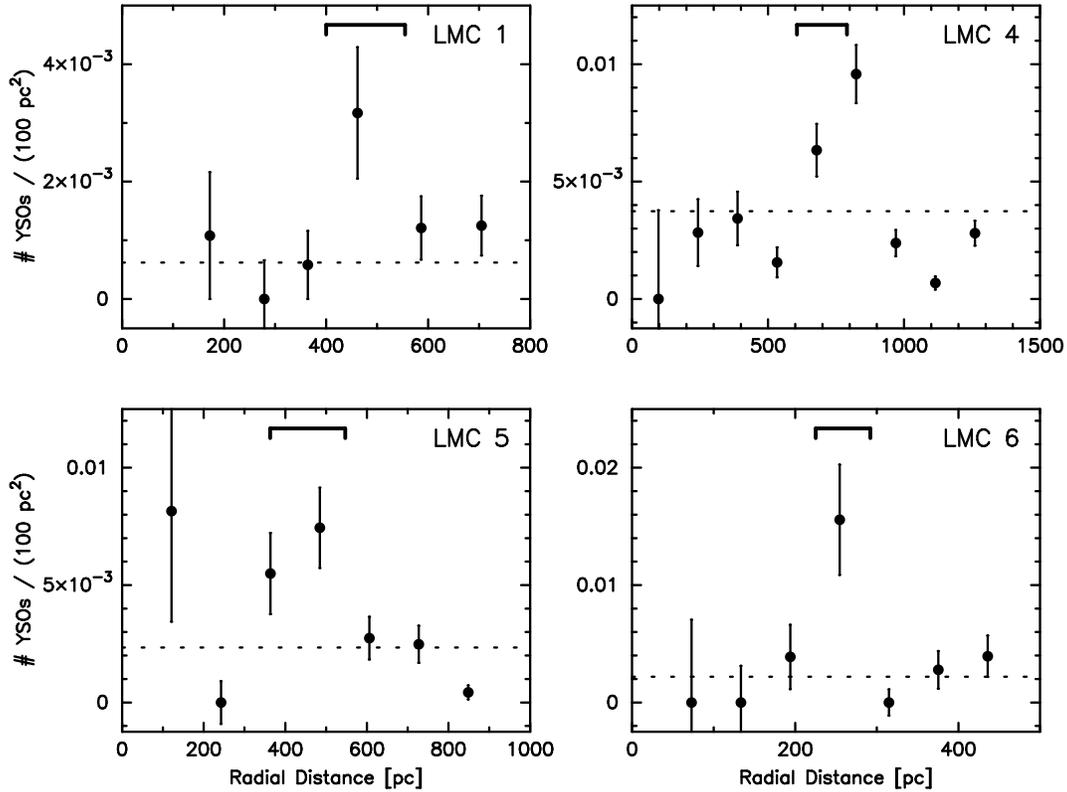}
\caption{Plot showing the surface density of YSOs 
in annular regions for each SGS.  The linear scale of each
plot gives the average radius of the trapezoidal regions used
for LMC\,1, and the semi-major axis of the elliptical annuli used
for LMC\,4, 5, and 6 (see text for details).  The rough extent
of the \HI\ shell is marked by a bracket near the top of each
panel.  The dashed horizontal line indicates the average surface
density of YSOs for the entire LMC at the galactocentric radius 
of the SGS center.}
\label{fig:sfquant}
\end{figure}

The resulting surface density of YSOs for each SGS is plotted in 
Figure~\ref{fig:sfquant}, which shows a clear peak along the rim of 
each SGS.   Furthermore, the surface density of YSOs along the SGS rims
is generally two to three times higher than the average for the LMC at 
the same galactic radius, similar to the results of \citet{Yetal01b}.
To show the statistical significance of the 
enhancements, we have plotted in Figure~\ref{fig:sfquant} error bars
that reflect the counting statistics in each annular region.   
The enhancement at the shell rim seen in LMC\,1, 5, and 6 suffer 
from small number statistics and are dominated by a few YSOs in a few 
relatively small areas.  These results may reflect a degree of 
organization in the star formation, but the uncertainty is large.
On the other hand, the number of YSOs in LMC\,4 that contribute to the 
pattern is much larger and do appear to be significantly higher than would
be expected.  Thus the amount of star formation on the periphery of LMC\,4 
appears to be statistically elevated, suggesting that the amount of star
formation is either augmented or organized by the SGS.

\subsection{Global Gravitational Instability}  \label{sec:grav}

Large-scale star formation in a galaxy is commonly believed to be driven
by global gravitational instability.  To investigate this hypothesis for
the LMC, \citet{Yang07} have considered a disk of gas only and a disk
of both collisional gas and collisionless stars. They find that
only 62\% of massive YSOs are in gravitationally unstable regions for the
former case and 85\% for the latter.  

It may be of interest to examine the relationship between the star
formation in SGSs and the global gravitational instability in the LMC.
We compare the locations of YSOs in the SGSs with Yang et al.'s (2007)
gravitational instability map that considered both the gas and stars.

The neutral shell of LMC 1 appears in the gravitational instability map 
to be almost entirely unstable, with only the northwestern corner closest 
to the edge of the LMC gravitationally stable (Fig.~4 of Yang et al.\
2007). All of the YSOs around LMC 1 are within the unstable region.

LMC 4 has an uneven distribution of unstable patches around its edge. While
most of the star formation is occurring in the unstable regions, there are
also a few YSOs located in regions that are gravitationally stable, such
as the few YSOs located in the interior of the shell.

LMC 5, like LMC 1, is composed of an almost entirely unstable rim, and all 
of the YSOs around it fall in unstable regions. Only two or three YSOs fall
in the interior of LMC 5, in a gravitationally stable region.

LMC 6 is located at the western end of the stellar bar of the LMC, a 
very dense region which is entirely gravitationally unstable. The 
shell itself is unstable all around the rim, with a stable cavity 
in the interior; all YSOs near LMC 6 are in the unstable regions.

These comparisons indicate that the great majority of YSOs formed
in SGSs LMC 1, 4, 5, and 6 are in gravitationally unstable regions,
where cloud collapse can be easily triggered by perturbations.
A small number of YSOs in LMC 1, 4, and 5 lie in gravitationally 
stable regions, and in all cases the YSOs are projected within the SGS
interiors and superposed on diffuse \HA\ emission. To investigate
the formation of these YSOs, the detailed local environments need
to be examined.

YSOs projected within SGS interiors have interstellar environments
similar to those of YSOs projected in superbubble interiors.
In the case of the superbubble N51D \citep{Henize56}, 
its interior YSOs have been
serendipitously imaged by the {\it Hubble Space Telescope} ({\it HST})
and shown to be embedded in dust globules whose surfaces are
photoionized by the OB associations in N51D \citep{Chu05}.  
The photoionized surface has a much higher thermal pressure than
the interior of the globule, and the implosion leads to star
formation.  It is possible that the YSOs in the gravitationally stable
interior of the SGSs are also formed in the same fashion.  
High-resolution {\it HST} images are needed to reveal the YSOs'
immediate environment and verify the existence of dust globules.

\section{Conclusion}  \label{sec:conc}

We have examined the distribution of YSOs and interstellar gas around 
the four simple coherent expanding \HA-selected SGSs LMC 1, 4, 5, and 6.
While we find different patterns and histories of star formation in 
these SGSs, we also find common characteristics in their current star
formation.

On the largest scale, recent star formation occurs preferentially
along the peripheries of the SGSs.  There is an excellent correlation
between young OB associations that are still in \HII\ regions and
the existence of molecular clouds.  These regions are also hosts of
higher concentration of massive YSOs, indicating that star formation
continues. 

Both collect-and-collapse and compression mechanisms may take place 
in the formation of molecular clouds.  LMC 1 has collected but not
collapsed; consequently, no active star formation is present in LMC 1.
LMC 4 and 5 have formed molecular clouds by both collect-and-collapse 
and compression, and active star formation (both OB associations and
massive YSOs) is seen along their rims, especially along their
interaction zone.  LMC 6 may have compressed preexisting molecular 
clouds and triggered the formation of two OB associations.

A small number of isolated YSOs are also seen along shell 
peripheries and even in shell interiors without detectable 
molecular clouds. Close examination shows that the formation of
most of these YSOs is triggered by local ionization fronts
advancing into small dark clouds.  Such local triggering of
star formation is important even in \HII\ regions where massive
stars have dynamically altered the physical structures of the 
interstellar gas.

Using the star formation in the SGSs LMC 1, 4, 5, and 6, we
have illustrated that the global gravitational instabilities
may have prepared the conditions for star formation, but the actual
star formation is begun by triggering, both on large scales
and local scales.

\acknowledgements
This research was supported by NASA grants JPL1264494 and JPL1290956.

%
%

\bibliographystyle{apj} 
\bibliography{ref}

\begin{thebibliography}{19}
\expandafter\ifx\csname natexlab\endcsname\relax\def\natexlab#1{#1}\fi

\bibitem[Bica et al.(1996)]{Betal96} Bica, E., Claria, J.~J., 
Dottori, H., Santos, J.~F.~C., Jr., \& Piatti, A.~E.\ 1996, \apjs, 102, 57 

\bibitem[Chen et al.(2008)]{Cetal08}
Chen, C.-H.\ R., Chu, Y.-H., Gruendl, R.~A., Gordon, K.~D., \& Heitsch, F.\
2008, submitted to \apj

\bibitem[{{Chu} {et~al.}(2005){Chu}, {Gruendl}, {Chen}, {Whitney}, {Gordon},
  {Looney}, {Clayton}, {Dickel}, {Dunne}, {Points}, {Smith}, \&
  {Williams}}]{Chu05}
{Chu}, Y.-H., et al.\ 2005, \apjl, 634, L189

\bibitem[{{Davies} {et~al.}(1976){Davies}, {Elliott}, \& {Meaburn}}]{DEM76}
{Davies}, R.~D., {Elliott}, K.~H., \& {Meaburn}, J.\ 1976, \memras, 81, 89

\bibitem[{{Domg\"orgen} {et~al.}(1995){Domg\"orgen}, {Bomans}, \& {de
  Boer}}]{Dom95}
{Domg\"orgen}, H., {Bomans}, D.~J., \& {de Boer}, K.~S.\ 1995, \aap, 296, 523

\bibitem[{{Elmegreen}(1998)}]{Elm98}
{Elmegreen}, B.~G.\ 1998, in Astronomical Society of the Pacific Conference
  Series, Vol. 148, Origins, ed. C.~E. {Woodward}, J.~M. {Shull}, \& H.~A.\
  {Thronson}, Jr., 150

\bibitem[{{Fazio} {et~al.}(2004){Fazio}, {Hora}, {Allen}, {Ashby}, {Barmby},
  {Deutsch}, {Huang}, {Kleiner}, {Marengo}, {Megeath}, {Melnick}, {Pahre},
  {Patten}, {Polizotti}, {Smith}, {Taylor}, {Wang}, {Willner}, {Hoffmann},
  {Pipher}, {Forrest}, {McMurty}, {McCreight}, {McKelvey}, {McMurray}, {Koch},
  {Moseley}, {Arendt}, {Mentzell}, {Marx}, {Losch}, {Mayman}, {Eichhorn},
  {Krebs}, {Jhabvala}, {Gezari}, {Fixsen}, {Flores}, {Shakoorzadeh}, {Jungo},
  {Hakun}, {Workman}, {Karpati}, {Kichak}, {Whitley}, {Mann}, {Tollestrup},
  {Eisenhardt}, {Stern}, {Gorjian}, {Bhattacharya}, {Carey}, {Nelson},
  {Glaccum}, {Lacy}, {Lowrance}, {Laine}, {Reach}, {Stauffer}, {Surace},
  {Wilson}, {Wright}, {Hoffman}, {Domingo}, \& {Cohen}}]{Fazio04}
{Fazio}, G.~G., et al.\ 2004, \apjs, 154, 10

\bibitem[{{Feast}(1999)}]{Feast99}
{Feast}, M.\ 1999, in IAU Symposium, Vol. 190, New Views of the Magellanic
  Clouds, ed. Y.-H. {Chu}, N.~{Suntzeff}, J.~{Hesser}, \& D.~{Bohlender},
  542

\bibitem[{{Fukui} {et~al.}(2001){Fukui}, {Mizuno}, {Yamaguchi}, {Mizuno}, \&
  {Onishi}}]{Fukui01}
{Fukui}, Y., {Mizuno}, N., {Yamaguchi}, R., {Mizuno}, A., \& {Onishi}, T.\ 2001,
  \pasj, 53, L41

\bibitem[{{Fukui} {et~al.}(1999){Fukui}, {Mizuno}, {Yamaguchi}, {Mizuno},
  {Onishi}, {Ogawa}, {Yonekura}, {Kawamura}, {Tachihara}, {Xiao}, {Yamaguchi},
  {Hara}, {Hayakawa}, {Kato}, {Abe}, {Saito}, {Mano}, {Matsunaga}, {Mine},
  {Moriguchi}, {Aoyama}, {Asayama}, {Yoshikawa}, \& {Rubio}}]{F99}
{Fukui}, Y., et al.\ 1999, \pasj, 51, 745

\bibitem[{{Gruendl} \& {Chu}(2008)}]{GC08}
{Gruendl}, R.~A. \& {Chu}, Y.-H.\ 2008, submitted to \apjs

\bibitem[{{Henize}(1956)}]{Henize56}
{Henize}, K.~G.\ 1956, \apjs, 2, 315 

\bibitem[{{Kim} {et~al.}(1999){Kim}, {Dopita}, {Staveley-Smith}, \&
  {Bessell}}]{Kim99}
{Kim}, S., {Dopita}, M.~A., {Staveley-Smith}, L., \& {Bessell}, M.~S.\ 1999,
  \aj, 118, 2797

\bibitem[{{Kim} {et~al.}(1998){Kim}, {Staveley-Smith}, {Dopita}, {Freeman},
  {Sault}, {Kesteven}, \& {McConnell}}]{Kim98}
{Kim}, S., et al.\ 1998, \apj, 503, 674

\bibitem[{{Kim} {et~al.}(2003){Kim}, {Staveley-Smith}, {Dopita}, {Sault},
  {Freeman}, {Lee}, \& {Chu}}]{Kim03}
{Kim}, S., et al.\ 2003, \apjs, 148, 473

\bibitem[{{Lucke} \& {Hodge}(1970)}]{lh}
{Lucke}, P.~B.\ \& {Hodge}, P.~W.\ 1970, \aj, 75, 171

\bibitem[{{Meaburn}(1980)}]{M80}
{Meaburn}, J.\ 1980, \mnras, 192, 365

\bibitem[{{Meixner} {et~al.}(2006){Meixner}, {Gordon}, {Indebetouw}, {Hora},
  {Whitney}, {Blum}, {Reach}, {Bernard}, {Meade}, {Babler}, {Engelbracht},
  {For}, {Misselt}, {Vijh}, {Leitherer}, {Cohen}, {Churchwell}, {Boulanger},
  {Frogel}, {Fukui}, {Gallagher}, {Gorjian}, {Harris}, {Kelly}, {Kawamura},
  {Kim}, {Latter}, {Madden}, {Markwick-Kemper}, {Mizuno}, {Mizuno}, {Mould},
  {Nota}, {Oey}, {Olsen}, {Onishi}, {Paladini}, {Panagia}, {Perez-Gonzalez},
  {Shibai}, {Sato}, {Smith}, {Staveley-Smith}, {Tielens}, {Ueta}, {Dyk},
  {Volk}, {Werner}, \& {Zaritsky}}]{Meixner06}
{Meixner}, M., et al.\ 2006, \aj, 132, 2268

\bibitem[{{Rieke} {et~al.}(2004){Rieke}, {Young}, {Engelbracht}, {Kelly},
  {Low}, {Haller}, {Beeman}, {Gordon}, {Stansberry}, {Misselt}, {Cadien},
  {Morrison}, {Rivlis}, {Latter}, {Noriega-Crespo}, {Padgett}, {Stapelfeldt},
  {Hines}, {Egami}, {Muzerolle}, {Alonso-Herrero}, {Blaylock}, {Dole}, {Hinz},
  {Le Floc'h}, {Papovich}, {P{\'e}rez-Gonz{\'a}lez}, {Smith}, {Su}, {Bennett},
  {Frayer}, {Henderson}, {Lu}, {Masci}, {Pesenson}, {Rebull}, {Rho}, {Keene},
  {Stolovy}, {Wachter}, {Wheaton}, {Werner}, \& {Richards}}]{Rieke04}
{Rieke}, G.~H., et al.\ 2004, \apjs, 154, 25

\bibitem[{{Smith} \& {The MCELS Team}(1999)}]{MCELS}
{Smith}, R.~C. \& {The MCELS Team}.\ 1999, in IAU Symp. 190: New Views of the
  Magellanic Clouds, ed. Y.-H. {Chu}, N.~{Suntzeff}, J.~{Hesser}, \&
  D.~{Bohlender}, 28

\bibitem[Walborn \& Parker(1992)]{WP92} 
Walborn, N.~R., \& Parker, J.~W.\ 1992, \apjl, 399, L87

\bibitem[{{Yamaguchi} {et~al.}(2001{\natexlab{a}}){Yamaguchi}, {Mizuno},
  {Onishi}, {Mizuno}, \& {Fukui}}]{Yetal01a}
{Yamaguchi}, R., {Mizuno}, N., {Onishi}, T., {Mizuno}, A., \& {Fukui}, Y.\
  2001{\natexlab{a}}, \apjl, 553, L185

\bibitem[{{Yamaguchi} {et~al.}(2001{\natexlab{b}}){Yamaguchi}, {Mizuno},
  {Onishi}, {Mizuno}, \& {Fukui}}]{Yetal01b}
---. 2001{\natexlab{b}}, \pasj, 53, 959

\bibitem[{{Yang} {et~al.}(2007){Yang}, {Gruendl}, {Chu}, {Mac Low}, \&
  {Fukui}}]{Yang07}
{Yang}, C.-C., {Gruendl}, R.~A., {Chu}, Y.-H., {Mac Low}, M.-M., \& {Fukui}, Y.\
  2007, \apj, 671,374

\bibitem[Zaritsky et al.(2004)]{Zetal04} Zaritsky, D., Harris, 
J., Thompson, I.~B., \& Grebel, E.~K.\ 2004, \aj, 128, 1606 

\end{thebibliography}

\end{document}